\documentclass[runningheads]{llncs}
\usepackage[T1]{fontenc}
\usepackage{amsmath}
\usepackage{graphicx}
\usepackage{hyperref}
\begin{document}
\title{A Deep User Interface for Exploring LLaMa}

%
%
\author{Divya Perumal \& Swaroop Panda}
%
\authorrunning{D Perumal \& S Panda}
%
\institute{\tt{Northumbria University}}
\maketitle              
\begin{abstract}
The growing popularity and widespread adoption of large language models (LLMs) necessitates the development of tools that enhance the effectiveness of user interactions with these models. Understanding the structures and functions of these models poses a significant challenge for users. Visual analytics-driven tools enables users to explore and compare, facilitating better decision-making. This paper presents a visual analytics-driven tool equipped with interactive controls for key hyperparameters, including top-p, frequency and presence penalty, enabling users to explore, examine and compare the outputs of LLMs. In a user study, we assessed the tool's effectiveness, which received favorable feedback for its visual design, with particular commendation for the interface layout and ease of navigation. Additionally, the feedback provided valuable insights for enhancing the effectiveness of Human-LLM interaction tools.

\end{abstract}

\section{Introduction}
The emergence of Large Language Models (LLMs), such as OpenAI's GPT, Google's BERT, and Meta's Llama, has fundamentally transformed the field of natural language processing. These models have introduced significant advancements in areas such as text generation \cite{li2024pre}, summarization \cite{zhang2024benchmarking}, and other related applications. In everyday tasks, these LLMs have also found widespread application, significantly enhancing user experiences across various domains. For instance, they are embedded in conversational voice assistants allowing users to interact with technology through natural, conversational language \cite{wang2023enabling,mahmood2023llm}.



Despite the impressive capabilities of LLMs, their complex nature often presents challenges in terms of interpretability and usability \cite{xu2024exploring}. As these models continue to grow in size and complexity, the need for effective tools to facilitate the communication and visualization of their inputs and outputs becomes increasingly urgent. For example, LLMs are being increasingly utilized in the medical field for diagnosing diseases \cite{chen2024rarebench}, recommending treatments \cite{wilhelm2023large}, and generating medical reports \cite{tan2024medchatzh}, as well as in data science, where they contribute to data analysis\cite{ma2023insightpilot}, trend prediction\cite{jin2024time}, and decision support \cite{wang2019human}. However, the opaque, \textit{black-box} nature of LLMs \cite{bhattacharjee2023llms,liu2018towards} presents a considerable barrier to validating the accuracy and reliability of their results. There remains a substantial gap in understanding \cite{liu2024understanding} how these models function and how they can be modified to produce different outcomes. The impact of these applications is largely contingent on the precision and appropriateness of the models, underscoring the critical need for transparency \cite{larsson2020transparency} and accountability \cite{doshi2017accountability} in AI systems, particularly as they become increasingly integrated into daily life.

Explainable AI and visual analytics \cite{alicioglu2022survey,epp2015uncertainty} methods have been adapted to mitigate this black box nature of LLMs.  These approaches facilitate a deeper understanding of model behavior by providing insights \cite{yi2008understanding} into decision-making processes and enabling users to visualize the relationships between input parameters and output predictions, thereby enhancing interpretability \cite{andrienko2022visual} and trustworthiness \cite{ooge2022explaining}. By employing techniques such as interactive visualizations\cite{godfrey2016interactive,batch2017interactive} or explorations\cite{jankun2007model}, these methods empower users to gain a clearer understanding of how LLMs generate responses. This increased transparency is essential for fostering user confidence\cite{kizilcec2016much} and ensuring responsible AI \cite{arrieta2020explainable} deployment, particularly in sensitive applications such as healthcare, finance, and legal decision-making.



In this paper,we present a visual analytics-driven tool to facilitate an understanding of working of LLMs by examining their underlying hyperparameters. The contributions of the paper are,
\begin{enumerate}
    \item We design and develop a deep user interface to facilitate the hyperparameter driven exploration of a Large Language Model
    \item We evaluate this user interface through user feedback and incorporate their suggestions to enhance its design.
    \item We provide preliminarily actionable insights for researchers to design and develop user interfaces for LLMs.
\end{enumerate}



\section{Background}
\subsection{Visual Analytics based User Interfaces}

Visual Analytics (VA) combines visualization, human interaction, and data analysis to support analytical reasoning through interactive visual interfaces \cite{cui2019visual}. Visual analytics tools such as model performance dashboards help in identifying issues like overfitting, bias, or data leakage by visually representing mode performance metrics enhance the AI interpretability \cite{yuan2021survey}. VA has also been used to understand the inner workings of neural networks by visualizing activations and weights help the interpretation of the model decisions \cite{garcia2018task}.  VA supports these cognitive processes by providing interactive tools that clear the way for deeper insights and better decision-making \cite{endert2017state}.




\subsection{Interpretability and Exploration for LLMs}

Interpretability in machine learning refers to the capacity to comprehend, trust, and manage a model's outputs and decision-making processes. The inherent black-box nature of LLMs, attributed to their deep neural network architectures comprising millions or even billions of parameters, presents significant challenges in this regard. Various interpretability methodologies have been developed, including post-hoc approaches, which generate explanations after a model has produced a prediction. Many of these methods do not directly interpret the model itself but rather attempt to elucidate its decision-making process. Several techniques, such as SHapley Additive exPlanations (SHAP) \cite{khediri2024enhancing} and Local Interpretable Model-agnostic Explanations (LIME) \cite{spinner2019explainer}, assign importance scores to input features. In the context of LLMs, these methods are commonly employed to analyze the significance of specific words, phrases, or tokens. The resulting visualizations aid in understanding how different linguistic elements influence predictions, thereby offering insights into complex model behavior. These techniques may not always capture the true reasoning of the model, potentially leading to misleading interpretations. Moreover, the application of these methods to LLMs, which involve extensive parameter sets and input tokens, is computationally expensive and complex. Many LLMs, such as BERT, utilize self-attention mechanisms to assign varying levels of importance to different words within a sentence \cite{insuasti2023computers}. Researchers have explored the interpretation of attention weights as proxies for understanding model decisions \cite{coscia2024iscore,la2023state}. However, the extent to which these weights provide meaningful insights into a model's reasoning remains an area of ongoing investigation.

\subsection{User Interfaces for LLMs}

One of the key challenges in utilizing LLMs is the selection of optimal hyper-parameters, which significantly influence model performance \cite{li2018hypertuner}. Visual analytics facilitates hyper-parameter tuning by providing interactive visualizations that help both experts and non-experts discern parameter relationships, enhancing model optimization. For example, Google Vizier employs parallel coordinates plots to analyze hyper-parameters, elucidating the relationships between inputs and outputs \cite{golovin2017google}. Sacha et al. \cite{sacha2017you} proposed an interactive machine learning framework that incorporates human feedback via visualization, improving parameter refinement. Similarly, Kahng et al. \cite{kahng2024llm} introduced a tool for comparing and evaluating LLM outputs during fine-tuning, addressing visualization challenges but still limited in scope and standardization. Chen et al. \cite{chen2024stugptviz} explored visual analytics in ChatGPT through StudentGPT, a tool that analyzes student interactions to derive cognitive insights. Despite the availability of such tools, there remains a gap in integrating visual analytics for hyper-parameter analysis where human intuition is crucial \cite{park2020hypertendril}.


Although visual analytics has been explored in various machine learning contexts, its application to hyper-parameter tuning in LLMs remains relatively under-explored, with limited dedicated tools. Hyperparameter tuning is crucial for optimizing model performance, yet it often remains an empirical process with limited interpretability. Visual analytics could aid in exploring the hyper-parameter space and understanding parameter interactions, though its impact on tuning efficiency requires further validation.

\section{Design \& Development of the User Interface}

The visual analytics tool was designed and developed to enable users to adjust model hyperparameters visually, making it easier to see how changes affect the LLM's outputs. This involved creating intuitive visualizations for each hyperparameter to facilitate user interaction. For this tool, we choose the open source large language model Llama \cite{touvron2023llama}. 

\subsection{Hyperparameters} 


We selected two hyperparameters, top p and frequency and presence penalty based on their availability via the LLaMA API.

\textbf{Top-P}: The top-p hyperparameter \cite{requeima2024llm}, also known as nucleus sampling, is a technique used in NLP to control text generation. It plays the role of selecting the next word in a sequence impacting the creativity of the generated text. In LLMs, content generation is the prediction of the next word in a sequence based on the words that generated previously. The model generates a list of all possible tokens and ranks them by their predicted probability. Top-p sampling is a method used to choose the next word based on these probabilities. The model accumulates the probabilities until the sum reaches a threshold value, $p$, which is set by the top-p parameter. At this point, the model forms a candidate pool of tokens, and the next token is randomly selected from this pool. This randomness introduces variation in the generated text, making it more creative. The top-p vocabulary $V^{(p)}$ is the smallest subset of the total vocabulary $V$  where the cumulative probability mass meets or exceeds the threshold value $p$ (Eq. \ref{eq}). The selection of the next word depends on ensuring that the cumulative probability is at least $p$, which adds randomness and diversity to the generated text. 


\begin{equation}
\sum_{x \in V^{(p)}} P(x \mid x_{1:i-1}) \geq p.
\label{eq}
\end{equation}


\textbf{Frequency and Presence penalty}: Frequency and presence penalty \cite{martinez2024beware} are two hyperparameters used in language models to control the repetition and diversity of the generated text. Both are designed to influence how often certain words or phrases appear in the generated text. Frequency penalty reduces the repetition of the same token within the generated text. The presence penalty encourages the model to introduce a new token into the generated text. The presence penalty works by reducing the probability of selecting a token that has already appeared in the text, regardless of how many times it has been used.  
To adjust token probabilities for diversity and repetition control, we define the new modified probability \( P'(t) \) for token \( t \) as:

\begin{equation}
P'(t) = \frac{P(t)}{(1 + \alpha \cdot f(t))(1 + \beta \cdot \mathbf{1}_{\{f(t) > 0\}})}
\end{equation}

where \( P(t) \) is the original probability, \( f(t) \) is the frequency of \( t \) in prior text, \( \alpha \) is the frequency penalty, and \( \beta \) is the presence penalty and \(\mathbf{1}_{\{f(t) > 0\}}\) is an indicator function that equals 1 if token \( t \) has appeared at least once in the text and 0 otherwise. The frequency penalty reduces \( P(t) \) proportional to \( f(t) \), while the presence penalty reduces \( P(t) \) if \( t \) has already appeared, enhancing diversity in the generated text.

\subsection{Development \& Visual Design}
The tool was developed using Python for backend logic and integration with the Llama API from \textit{Meta LLAMA-7B}, Flask for building the web interface, d3js for interactive data visualizations in web browsers and MongoDB for storing user interactions and survey responses. 



To visualize the top-p hyperparameter we use a knob (clock-like) interface that allows users to adjust the parameter values (which controls the nucleus sampling in the language model). This visual representation aims to provide an intuitive way for users to adjust and understand the impact of the top-p hyperparameter on text generation, allowing for more or less randomness in the output based on the setting.

To visualize the frequency and presence penalty we use a co-ordinate plane graph with x and y axis which represent presence and frequency penalty respectively. This scatter plot-type of visualization aids users in comprehending the relationship between these two hyperparameters as reflected in the model's output.

\begin{figure}[htb!]
\centerline{\includegraphics[width=1\linewidth]{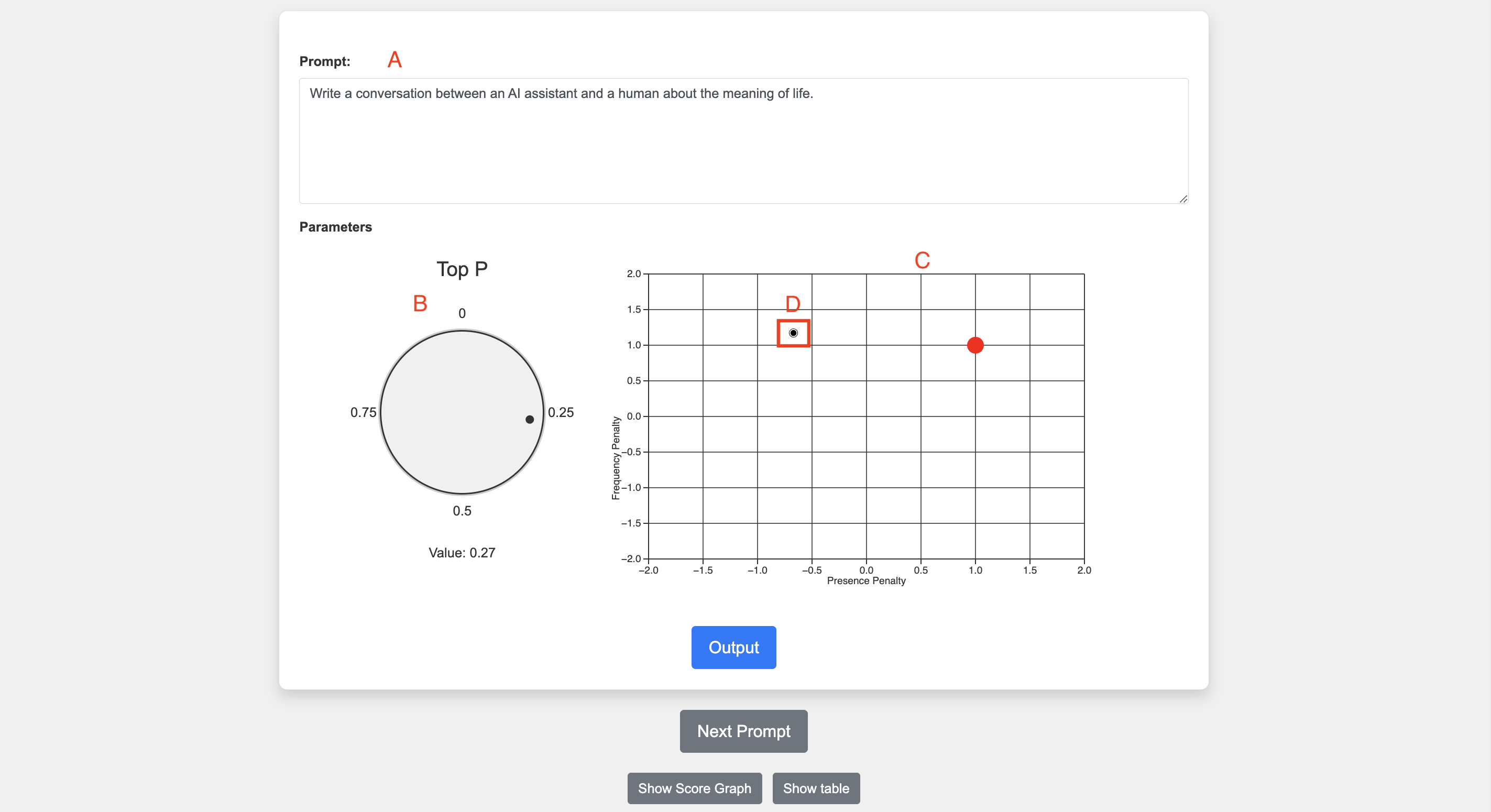}}
\caption{Visualization of Hyperparameters in the UI}
\label{tool1}
\end{figure}

In Fig.~\ref{tool1} (A) is the prompt section where the user enters their prompts to interact with the LLM. (B) represents the hyperparameters section, which is designed to be visually engaging. (C) is the graphical representation to adjust the presence and frequency penalty. (D) denotes the previous point used, which enables the user to compare outputs given the hyperparameter value.   

\begin{figure}[htb!]
\centerline{\includegraphics[width=1\linewidth]{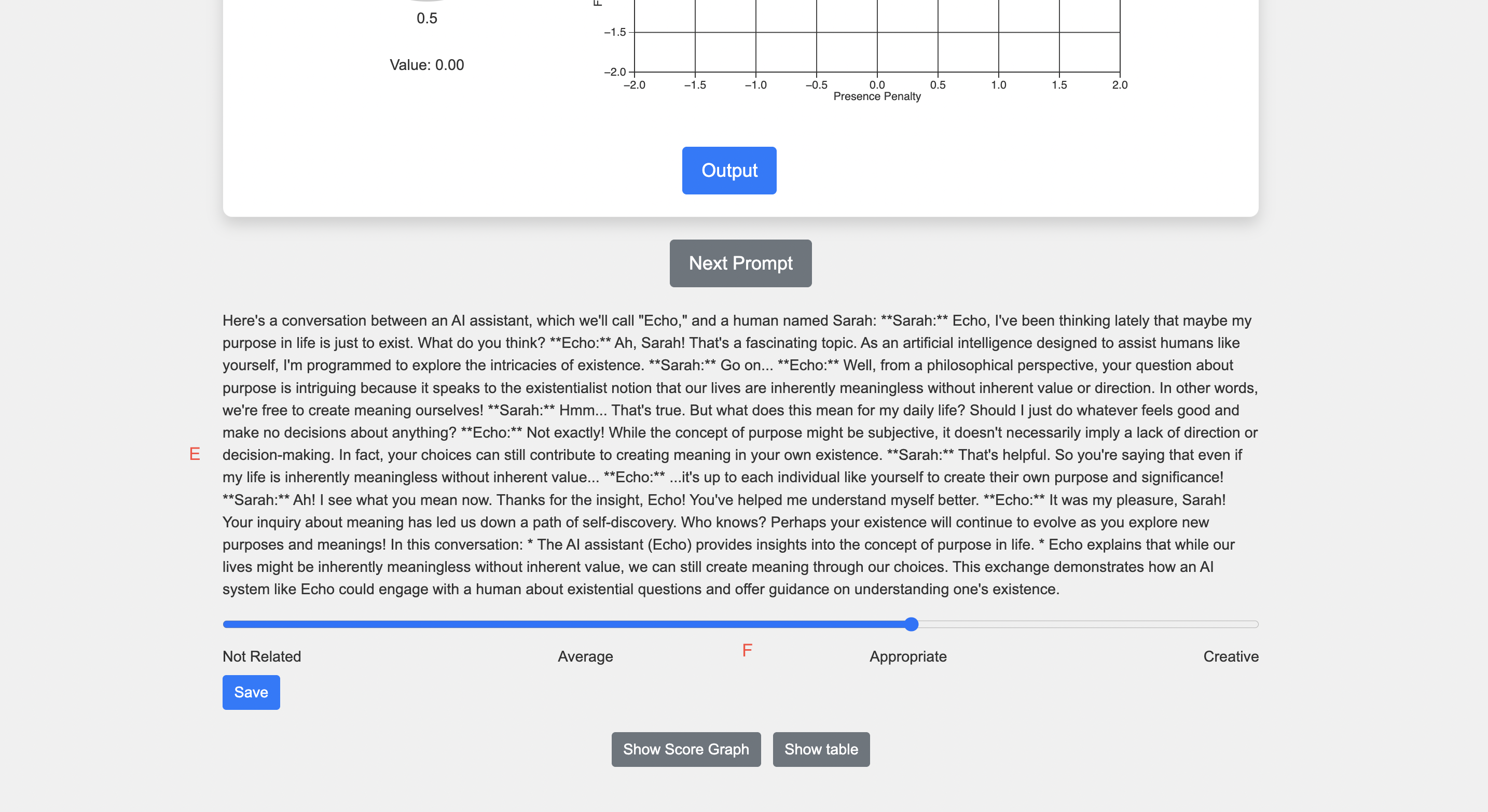}}
\caption{Visual Representation of LLM Outputs with the Scale for User Rating}
\label{tool2}
\end{figure}

Fig.~\ref{tool2} (E) denotes the output of the LLaMA. The content given in the image is for sample. (F) is the rating scale for the users to score the generated output; for the user study.



\begin{figure}[htb!]
\centerline{\includegraphics[width=1\linewidth]{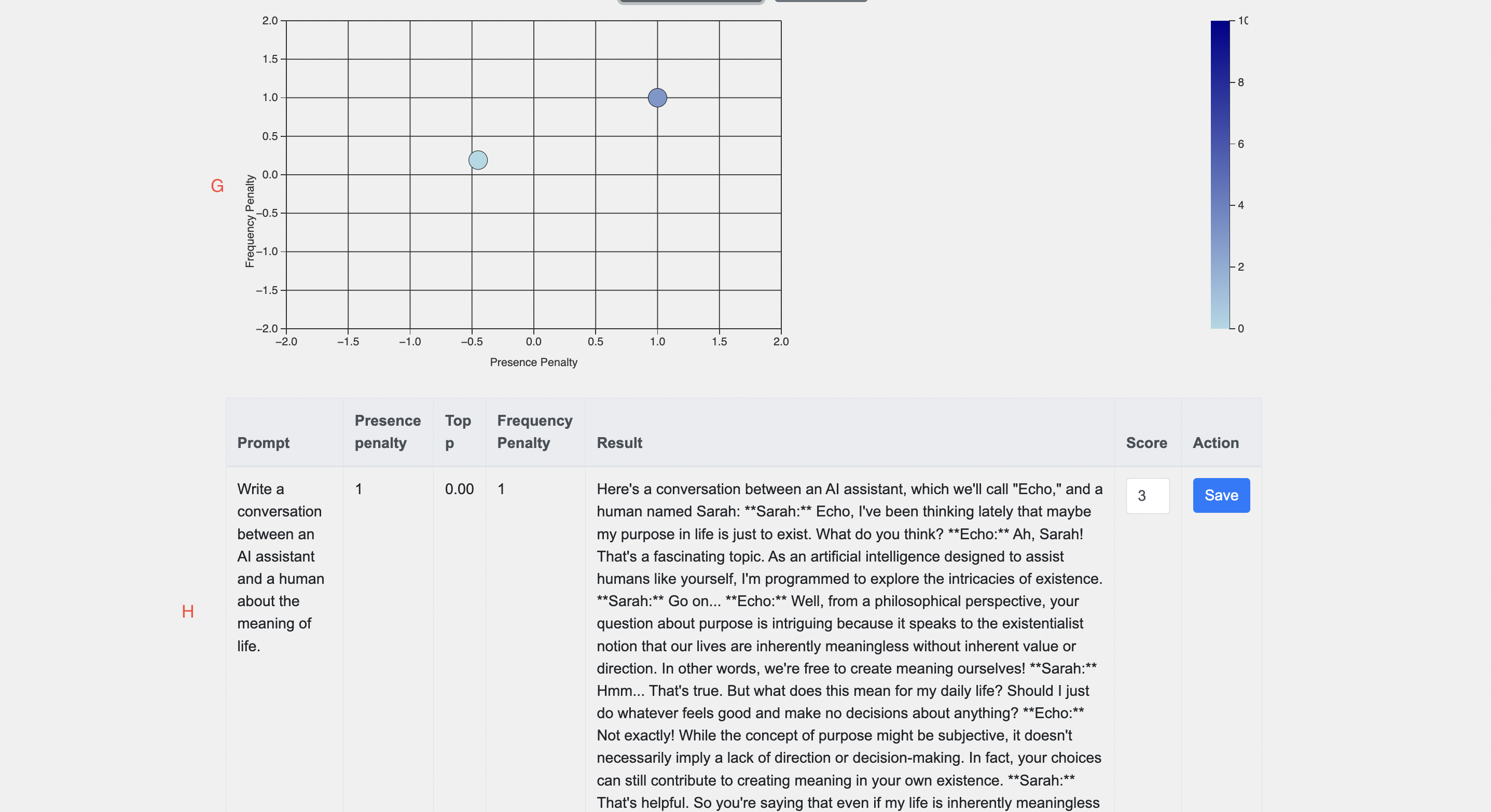}}
\caption{Visual Representation of \textbf{Score Graph} and \textbf{Table}}
\label{tool3}
\end{figure}

Fig.~\ref{tool3} (G) represents the graph with points of frequency and presence previously used for the given prompt (from Fig.~\ref{tool1} D). The shade of the points represents the score given to the output by the user. Fig.~\ref{tool3} (H) indicates the table with all the prompts, parameter values used and rating for each output.  

\section{Evaluation of the User Interface}
\subsection{User Study}
With approval from the relevant institutional ethics committee, participants were selected from among students enrolled in postgraduate programs, ensuring a decent understanding of LLM usage. Participants had academic backgrounds in computer and data science. A total of 10 participants took part in the experiment. Participants were of the age group \textit{mean:24.3 yrs, stddev:0.82}. 





Participants received a list of ten predefined prompts and were instructed to select three. They subsequently inputted these chosen prompts into the visual analytics tool. Utilizing the tool's visualized interfaces, participants adjusted hyperparameters— top-p, frequency penalty, and presence penalty—to examine their impacts on the model's outputs. After setting these hyperparameters, participants submitted the prompts to the LLaMA to generate corresponding text outputs. Each output was subsequently evaluated on a standardized rating scale.These tasks and exploration  were designed to assess how effectively participants could use the tool to change hyperparameters and compare, explore, evaluate the generated outputs. Followed by the experiment participants were asked to complete survey with combination of likert scale and qualitative questions to gain feedback on the tool's functionalities, interface and overall user experience. 


\subsection{Results \& Analysis}

Based on the user study results, participants provided overall positive feedback, particularly regarding the tool’s interface and its visual appeal (Table \ref{tab1}). Users found the interface visually engaging, with the layout receiving especially high ratings, indicating that the design is well-organized and easy to navigate.  The interface of the tool is found to be visually appealing to the users, with a mean rating of 4.22 and a relatively low standard deviation of 0.83, indicating the most respondents appreciated the design. The layout of the interface was rated even higher, with a mean of 4.56 and a standard deviation of 0.53, reflecting that the interface was well organized and easy to navigate.   

Qualitative feedback yielded valuable insights; users expressed a desire for more clarification regarding the functionality of the hyperparameters. Participants highlighted features that enhanced their experience with the system. They appreciated the ability to see varying outputs for prompts. One participant stated that \textit{"It was a good experience to see varying outputs for the prompts. The outputs differed on the parameters which provided different kinds of outputs which could then be rated as creative, average, or satisfying."}. The feature of adjusting hyperparameters using visualizations stood out as particularly convenient, with a participant noting, "\textit{the feature of adjusting parameters using visualizations was the most convenient one, along with the option of accessing history, which helped to compare the previous parameters."} Access to history was also valued for its role in facilitating comparisons between different parameter settings. The user interface was praised for its simplicity, making the selection of parameters straightforward. One participant suggested \textit{"selecting the parameters was easy as it was simple as the user interface was simple. For the prompts provided, it was helpful to play around with the parameters to check various outputs."}. Another participant stated, \textit{"most helpful parameter my opinion would be changing the answers as user changes the parameter."} 

Following the user study, we refined the tool based on user feedback to enhance clarity and usability. One key improvement was the addition of written descriptions for each hyperparameter, providing users with a clearer understanding of their influence on the model’s behavior. This change aims to make the tool more accessible, especially for those with limited prior knowledge, enabling more informed adjustments and interactions.

\begin{table}[htbp]
\begin{center}
\begin{tabular}{|p{9cm}|p{1cm}|p{1.5cm}|}
\hline
\textbf{\textit{Question }} & \textbf{\textit{Mean}} & \textbf{\textit{StdDev}} \\
\hline
How visually appealing do you find 
the interface of the visual analytics tool?  & 4.22 & 0.83  \\
\hline
How well-organized is the layout of the tool's interface?  & 4.55 & 0.52  \\

\hline
How readable and appropriate is the typography used in the tool?  & 4.66  & 0.5   \\

\hline
How easy was it for you to navigate and use the visual analytics tool?  & 4.33  & 0.86  \\

\hline
How effective are the tool’s features in helping you achieve your goals?  & 4.33  & 0.70  \\

\hline
How effective is the answer changing according to the given parameters?  & 4.22  & 0.66   \\

\hline
How clear and understandable are the visualizations provided by the tool?  & 4.33  & 0.5  \\

\hline
How would you rate the interactivity of the visualizations?  &4.3 & 0.67  \\

\hline
\end{tabular}
\caption{Results of the User Study on a 5-point Likert Scale}
\label{tab1}
\end{center}
\end{table}

\section{Discussion}
While the findings of the user study offer valuable perspectives, it is important to interpret them with caution due to the limited sample size. While the tool's navigation was generally rated as easy with a mean score of 4.33, the standard deviation of 0.86 suggests that a minority of users encountered difficulties. The tool's functional effectiveness, particularly in how well the features helped users achieve their goals, the responses were favorable. This indicates that the tool was generally effective though there were some variations in user experience, suggesting that certain features might benefit from further enhancement to ensure consistent effectiveness.

\subsection{Actionable Insights}
Preliminary insights suggest some considerations for the design of user interfaces for LLMs. These following actionable insights provide guidance on optimizing interaction paradigms, improving usability, and enhancing the overall user experience.

\begin{enumerate}
    \item Providing users with access to adjustable hyperparameters within the interface appears to facilitate exploration, particularly when these hyperparameters have a substantial impact on the generated outputs. This feature may enhance user engagement and comprehension of model behavior.
    \item While enabling hyperparameter-based exploration, it is beneficial to incorporate a history of generated responses. Such a feature allows users to compare outputs systematically, fostering a clearer understanding of how hyperparameter adjustments influence the model’s responses over time.
\end{enumerate}
\subsection{Future Work}
Future research includes exploring more hyperparameters of LLMs and developing visualizations that enhance user understanding, thereby facilitating more targeted LLM outputs. Furthermore, a more comprehensive user study involving diverse participants could reveal patterns in hyperparameter adjustments and user ratings, thereby providing deeper insights into user behavior and preferences.

\section{Conclusion}

The aim of our study was to explore how visual analytics could be integrated into the process of tuning hyperparameters in LLMs to improve user experience and make these models more interpretable. From our initial findings, we discovered that a more intuitive user interface—one that lets users adjust model settings, view and compare results—can transform complex models like LLama from a "black box" into something more transparent and user-friendly. The positive feedback we received on the design of this tool shows great potential for its wider use, making these advanced models more accessible and easier to understand.
\newpage

\bibliographystyle{splncs04}
\bibliography{refs}
%




\end{document}